\documentclass[usenatbib,useAMS,twocolumn]{aastex61}
\usepackage{epsfig,amssymb,amsmath,rotating,lineno,graphicx}
\newcommand\swift{{\it SWIFT}}

\newcommand\igl{{\it INTEGRAL}}

\newcommand\xte{{\it RXTE}}
\newcommand\asat{{\it AstroSat}}

\submitjournal{ApJ}
\shorttitle{LAXPC observations of Cyg X-3}
\shortauthors{Pahari et al.}
\begin{document}
\title{X-ray timing analysis of Cyg X-3 using AstroSat/LAXPC: Detection of milli-hertz quasi-periodic oscillations during the flaring hard X-ray state}
\correspondingauthor{Mayukh Pahari}
\email{mayukh@iucaa.in}

\author{Mayukh Pahari}
\affiliation{Inter-University Center for Astronomy and Astrophysics, Ganeshkhind, Pune 411007, India}

\author{H M Antia}
\affiliation{Tata Institute of Fundamental Research, Homi Bhabha Road, Mumbai 400005, India}

\author{J S Yadav}
\affiliation{Tata Institute of Fundamental Research, Homi Bhabha Road, Mumbai 400005, India}

\author{Jai Verdhan Chauhan}
\affiliation{Tata Institute of Fundamental Research, Homi Bhabha Road, Mumbai 400005, India}

\author{P C Agrawal}
\affiliation{UM-DAE Center of Excellence for Basic Sciences, University of Mumbai, Kalina, Mumbai 400098, India}

\author{Ranjeev Misra}
\affiliation{Inter-University Center for Astronomy and Astrophysics, Ganeshkhind, Pune 411007, India}

\author{V R Chitnis}
\affiliation{Tata Institute of Fundamental Research, Homi Bhabha Road, Mumbai 400005, India}

\author{Dhiraj Dedhia}
\affiliation{Tata Institute of Fundamental Research, Homi Bhabha Road, Mumbai 400005, India}

\author{Tilak Katoch}
\affiliation{Tata Institute of Fundamental Research, Homi Bhabha Road, Mumbai 400005, India}

\author{P Madhwani}
\affiliation{Tata Institute of Fundamental Research, Homi Bhabha Road, Mumbai 400005, India}

\author{R K Manchanda}
\affiliation{University of Mumbai, Kalina, Mumbai 400098, India}

\author{B Paul}
\affiliation{Department of Astronomy and Astrophysics, Raman Research Institute, Bengaluru 560080, India}

\author{Parag Shah}
\affiliation{Tata Institute of Fundamental Research, Homi Bhabha Road, Mumbai 400005, India}

\label{firstpage}

\begin{abstract}

We present here results from the X-ray timing and spectral analysis of the X-ray binary Cyg X-3 using observations from Large Area X-ray proportional Counter (LAXPC) on-board \asat{}. Consecutive lightcurves observed over a period of one year show the binary orbital period of 17253.56 $\pm$ 0.19 sec. Another low-amplitude, slow periodicity of the order of 35.8 $\pm$ 1.4 days is observed which may be due to the orbital precession as suggested earlier by \citet{mo80}. During the rising binary phase, power density spectra from different observations during flaring hard X-ray state show quasi-periodic oscillations (QPOs) at $\sim$5-8 mHz, $\sim$12-14 mHz, $\sim$18-24 mHz frequencies at the minimum confidence of 99\%. However, during the consecutive binary decay phase, no QPO is detected up to 2$\sigma$ significance. Energy-dependent time-lag spectra show soft lag (soft photons lag hard photons) at the mHz QPO frequency and the fractional rms of the QPO increases with the photon energy. During the binary motion, the observation of mHz QPOs during the rising phase of the flaring hard state may be linked to the increase in the supply of the accreting material in the disk and corona via stellar wind from the companion star. During the decay phase the compact source moves in the outer wind region causing the decrease in supply of material for accretion. This may cause weakening of the mHz QPOs below the detection limit. This is also consistent with the preliminary analysis of the orbital phase-resolved energy spectra presented in this paper.      
    
\end{abstract}


\section{Introduction}

Discovered nearly 40 years ago \citep{gi67}, Cygnus X-3 is one of the brightest, persistent and extra-ordinary Galactic high mass X-ray binaries. The X-ray emission is driven by the wind accretion from the massive Wolf-Rayet companion star ($\sim$ 20 M$\odot$; \citet{fe99,sz08,vi09}) along with relatively bright persistent radio emission virtually all of the time ($\sim$100 mJy) and occasionally accompanied by major radio ejection events of the order 10 Jy or more \citep{mo88,sc95,mi01}. The source resides close to the Galactic plane at a distance of 8-10 kpc \citep{di83}. A strong X-ray flux modulation with 4.8 hours cycle is observed \citep{pa72,le75,ki89} and attributed to the binary orbital period of the system. The rate of change of the binary orbital period has been measured as $\sim$10$^{-9}$ s.s$^{-1}$ \citep{va81,ki89}. However, the binary orbital period derivative has been revised ($\sim$ 5 $\times$ 10$^{-10}$ s.s$^{-1}$) by \citet{si02} using multi-mission data spanning over seven years. Longer modulation in X-rays ($\sim$ 34.1 days; \citet{ho76, mo80}) and in radio lightcurve ($>$ 60 days) are thought to be connected with the precessional motion of the accretion disk \citep{mi01}. In spite of extensive study, little is known about the compact nature of the source. Based on few evidences and similarities with black hole systems like GRS 1915+105, XTE J1550-564, the compact object is suggested to be a black hole \citep{sz08a} which is yet to be confirmed. Linking {\it RXTE}/PCA observations to the simultaneous radio observations, \citet{ko10} studied hardness intensity diagram and found six different spectral states. In addition to canonical spectral states of black hole X-ray binaries, they identified a hyper-soft spectral state/radio quenched state when strong GeV flux is detected using Fermi telescope \citep{bo13}. 

The study of the X-ray spectra at different spectral states and its timing variability have suffered severely from the poor understanding of the properties of the surrounding medium. As a consequence, a detailed interpretation of the Cyg X-3 intrinsic unabsorbed spectrum and luminosity is still missing. Interestingly, \citet{ma02} found that number of X-ray photons in 2-500 keV energy range is conserved irrespective of the source spectral state and such behavior is explained using a accretion geometry in which a thermal X-ray source is surrounded by hot plasma formed by the wind from accretion disk. Based on simultaneous \igl{} and \xte{} spectral modelling, \citet{vi03} interpreted the physical nature of X-ray emission as the strongly absorbed thermal Comptonization including Compton reflection and with parameters similar to other X-ray binaries at high accretion rates. Similar efforts are also given (e.g., \citet{sz04,zd10}) to understand nature of different accretion states in this peculiar source.

Despite of the availability of large archival data from {\it RXTE}/PCA, the X-ray
timing properties of Cyg X-3 are not particularly well studied. One reason could be the absence of high frequency power in the PDS above 0.1 Hz (a result that has been discussed in \citet{be94} (they placed an upper limit of 12 per cent
rms above 1 Hz)) where most of other low mass X-ray binaries show low- and high-frequency QPOs. This is believed to be an effect of scattering in the nearby surrounding medium \citep{zd10}. The PDS from Cyg X-3 is well-described by a power-law with the index between -1.8 \citep{wi85} and -1.5 \citep{ch04}. QPOs from Cyg X-3 at mHz frequency range have been reported few times in the literature. {\it EXOSAT}/ME observed 0.7-15 mHz QPOs in the PDS during the soft state with rms amplitude between 5-15 per cent \citep{va85}. A balloon study \citep{ma93,ra91} showed a QPO in the 20-100 keV lightcurve with a frequency of 8 mHz (121 s) and a pulse fraction of 40 per cents. However, contrary to these results, a study based on \xte{} pointing data of Cyg X-3 from 1996-2000 \citep{ax09} found no evidence of QPOs on any time-scale. Specifically, on the shorter time-scales ($>$ 10$^{-3}$ Hz) no QPO is detected and the power density spectrum is well described by a power law of index -2 (which is typical to the red noise that essentially mimics a random walk process), while on the longer time-scales ($<$10$^{-3}$ Hz), the variability is found to be dominated by the state transitions. The transient and time-sensitive nature of such QPOs could be a reason of discrepancy between results supporting the detection and non-detection of QPOs.   

\begin{figure*}
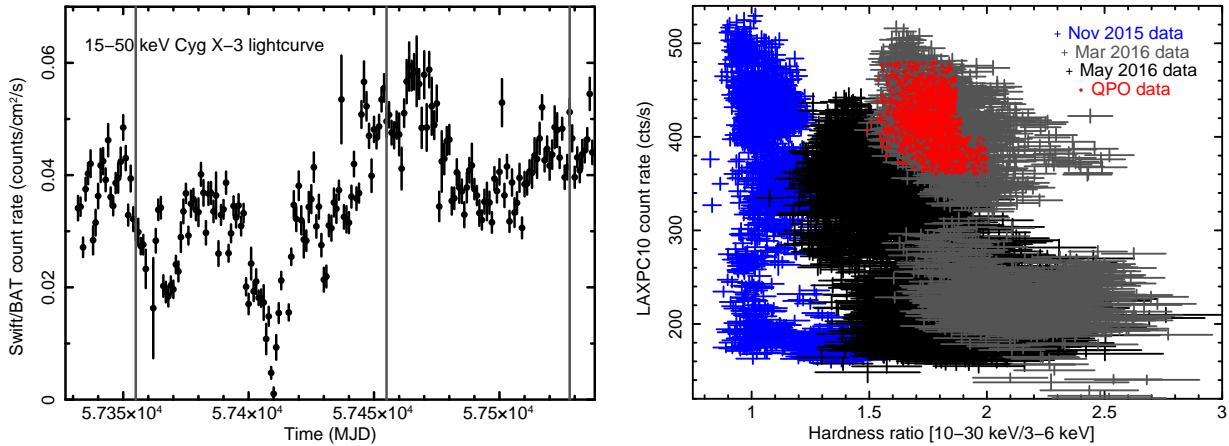

\includegraphics[scale=0.33,angle=-90]{fig1a.ps}
\includegraphics[scale=0.33,angle=-90]{fig1b.ps}
\caption{Left panel shows 15-50 keV \swift{}/BAT lightcurve of Cyg X-3 where three vertical, grey lines mark the epoch of \asat{}/LAXPC observations on November 2015, March 2016 and May 2016. The right panel shows the hardness intensity diagram (HID) where the 3-30 keV LAXPC10 count rate is shown as a function of hardness ratio (ratio of count rate between 10-30 keV and 3-6 keV). The HID of Nov 2015 observation (shown in blue symbols) is clearly softer than the HID of March 2016 (shown in grey symbols). However, the HID during May 2016 observation (shown in black symbols) occupy a region between Nov 2015 and March 2016 having a partial overlap with March 2016 HID. The HID position when mHz quasi-periodic oscillations are detected during March 2016 observation are shown by red symbols.   }
\label{hid}
\end{figure*}

\begin{table*}
 \centering
 \caption{\asat{}/LAXPC observation details of Cyg X-3. During all LAXPC observations, the source showed three states: flaring hard X-ray state (FHXR), flaring intermediate state (FIM) and flaring soft X-ray state (FSXR). }
\begin{center}
\scalebox{0.75}{%
\begin{tabular}{cccccccccc}
\hline 
Obs & orbit & Date & all-orbit combined & average source & spectral & QPO observed ? & QPO frequency & QPO & binary phase of\\
ID & number & (dd-mm-yyyy) &  exposure (sec) &  count rate & state & (99\% confidence) & (mHz) & occurrence & QPO intervals \\
\hline 
58 & 377-386 & 24-10-2015 & 17364 & 1329 $\pm$ 17 & FHXR & Yes & 13.6 $\pm$ 0.7 & 1 & 0.338-0.466\\
98 & 683-697 & 13-11-2015 & 36475 & 1389 $\pm$ 25 & FIM & No & -- & -- & -- \\
180 & 896-904 & 27-11-2015 & 18239 & 1529 $\pm$ 33 & FIM & No & -- & -- & -- \\
360 & 2375-2393 & 06-03-2016 & 34203 & 1432 $\pm$ 24 & FHXR & Yes & 20.3 $\pm$ 1.1 & 3 & 0.459-0.604\\
 &  &  &  &  &  &  &  &  & 0.176-0.367\\
 &  &  &  &  &  &  &  &  & 0.226-0.347\\
466 & 3500-3508 & 21-05-2016 & 25885 & 1494 $\pm$ 25 & FHXR & Yes & 5.7 $\pm$ 1.7 & 1 & 0.452-0.612\\
522 & 4094-4099 & 30-06-2016 & 15172 & 1405 $\pm$ 22 & FIM & No & -- & -- & -- \\
526 & 4110-4114 & 01-07-2016 & 11016 & 1544 $\pm$ 28 & FIM & No & -- & -- & -- \\
594 & 4727-4737 & 12-08-2016 & 27183 & 2433 $\pm$ 39 & FSXR & No & -- & -- & -- \\
812 & 6204-6218 & 20-11-2016 & 33788 & 1418 $\pm$ 23 & FHXR & Yes & 5.9 $\pm$ 1.9 & 2 & 0.378-0.552\\
 &  &  &  &  &  &  &  &  & 0.395-0.565\\
\hline
\end{tabular}}
\end{center}
\label{obs}
\end{table*}

\citet{ko11} re-analyzed archival \xte{} data of the X-ray binary Cyg X-3 and they identified two additional instances of quasi-periodic oscillations above the 99.9 per cent confidence limit, that have centroid frequencies in the mHz regime. The first one is detected at 8.5 mHz ($\sim$120 s) on 03 April, 2000, 2.2 days after the peak of a major radio flare ($\sim$13 Jy in the 15 GHz band) when the source was in the flaring soft X-ray state. During the flaring hard X-ray state observed on 9 August, 2009 a strong 21 mHz ($\sim$50 s) QPO was detected which corresponds to a 1.5-2 per cent rms in the PDS and lasts for 20 orbital cycles. The 15 GHz radio flux density during the second observation was $\sim$50 mJy. Based on simple arguments, they rejected the idea that QPOs are X-ray emitting blobs in a region of the accretion disk determined by the inner disk radius or the Roche Lobe radius.
Other possibilities like an oscillating corona due to a magneto-acoustic wave propagation within the corona, an oscillation in the jet-base caused by the downward propagation of relativistic shock from the jet, oscillation due to wind etc. have been discussed but none are confirmed as the origin of mHz QPOs. 

With the successful launch of first Indian multi-wavelength astronomical mission AstroSat in 2015 \citep{ag06,si14}, a new window to investigate spectro-temporal properties of Cyg X-3 has been opened. Large Area X-ray proportional Counter (LAXPC), one of the payloads in \asat{}, consists of three independent units of proportional counters ({\tt LAXPC10}, {\tt LAXPC20}, {\tt LAXPC30}) with the key feature of large effective area $\sim$4500 cm$^2$ at 30 keV ($\sim$4-5 time higher than that of RXTE/PCA; \citet{ya16a}) and absolute time resolution of 10$\mu$s. Details of the instrument and calibration can be found in \citet{ya16b}. LAXPC has already shown its capability to perform spectro-temporal variability study in micro-quasars like GRS 1915+105 \citep{ya16a} and Cyg X-1 \citep{mi16}.
In this work, using several orbit data of Cyg X-3 during several epochs spanning over $\sim$400 days we measure the binary orbital period of the system of 17253.56 $\pm$ 0.19 sec and reconfirm the detection of $\sim$20 mHz QPO during flaring hard X-ray state in Cyg X-3 using LAXPC on-board \asat{}. Energy-dependent study shows soft-lag at 20 mHz (soft photons lag hard photons). We also perform spectral analysis during the rising and decay phase of the binary orbit to understand the connection between energy spectral and QPO properties as a function of binary phase. Section 2 provides observation and analysis procedures while results are provided in section 3. Discussions and conclusions are provided in section 4.

\begin{figure*}
\includegraphics[width=8.5cm,height=5.6cm]{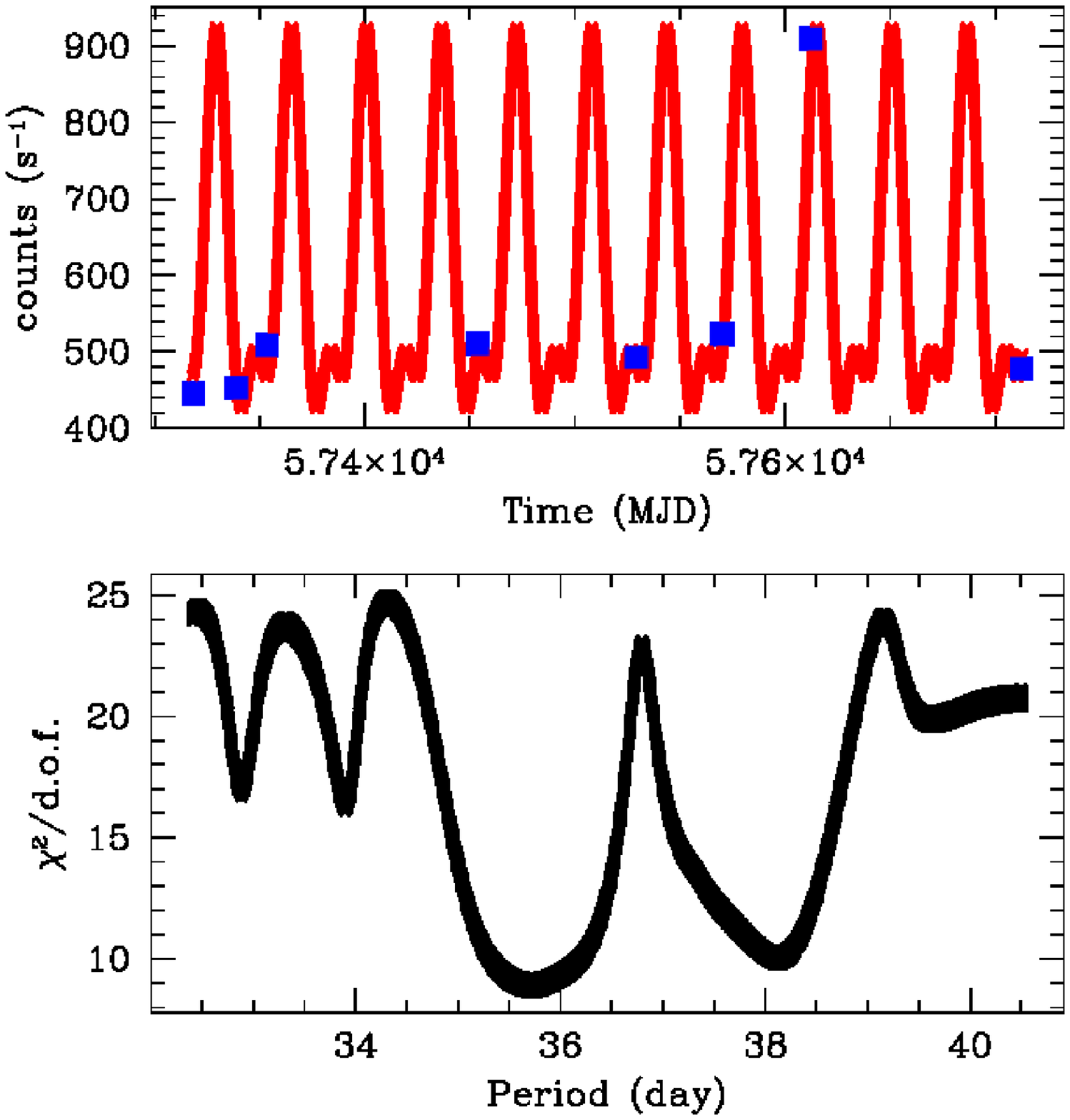}
\includegraphics[scale=0.245]{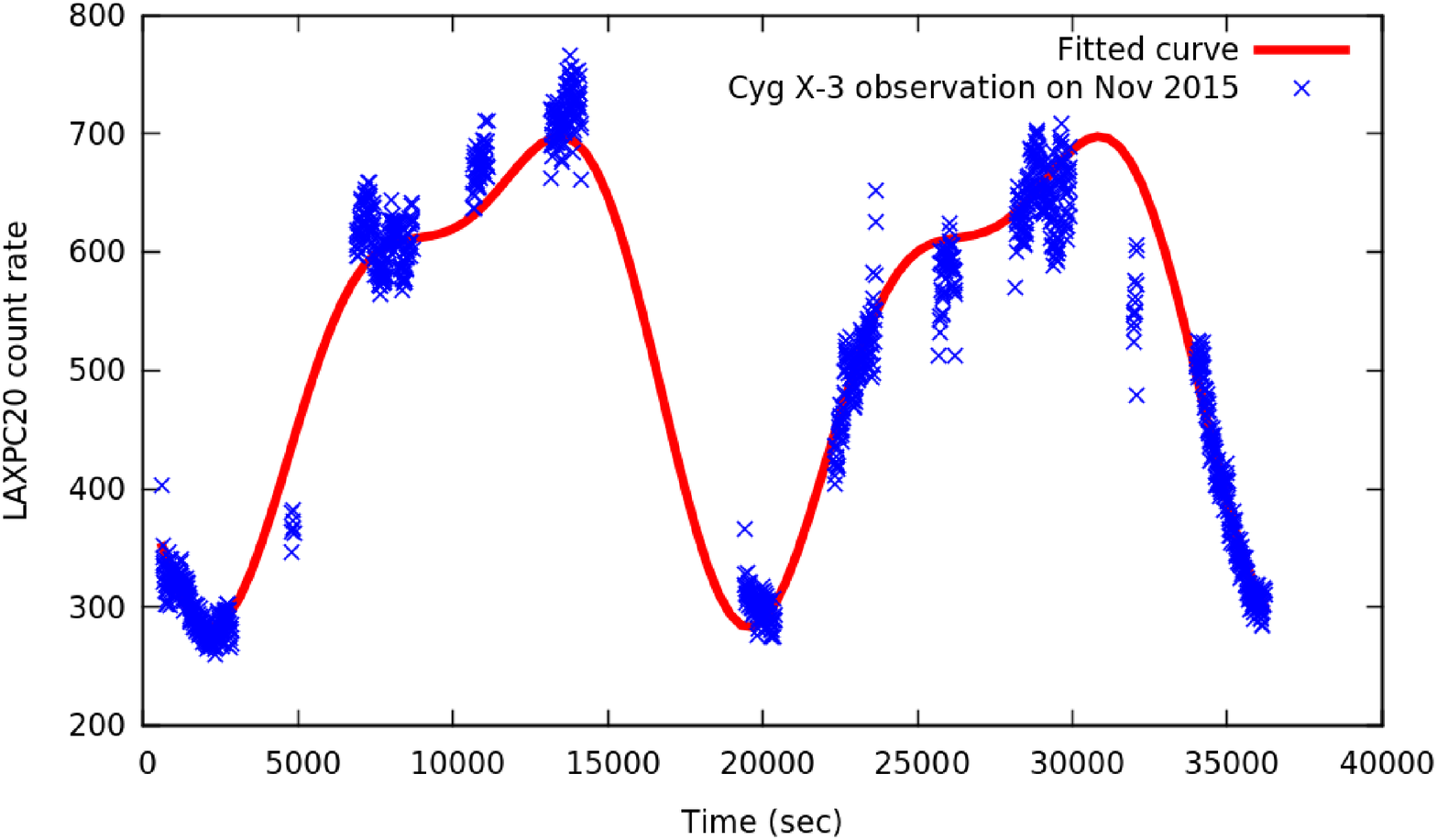}
\includegraphics[scale=0.245]{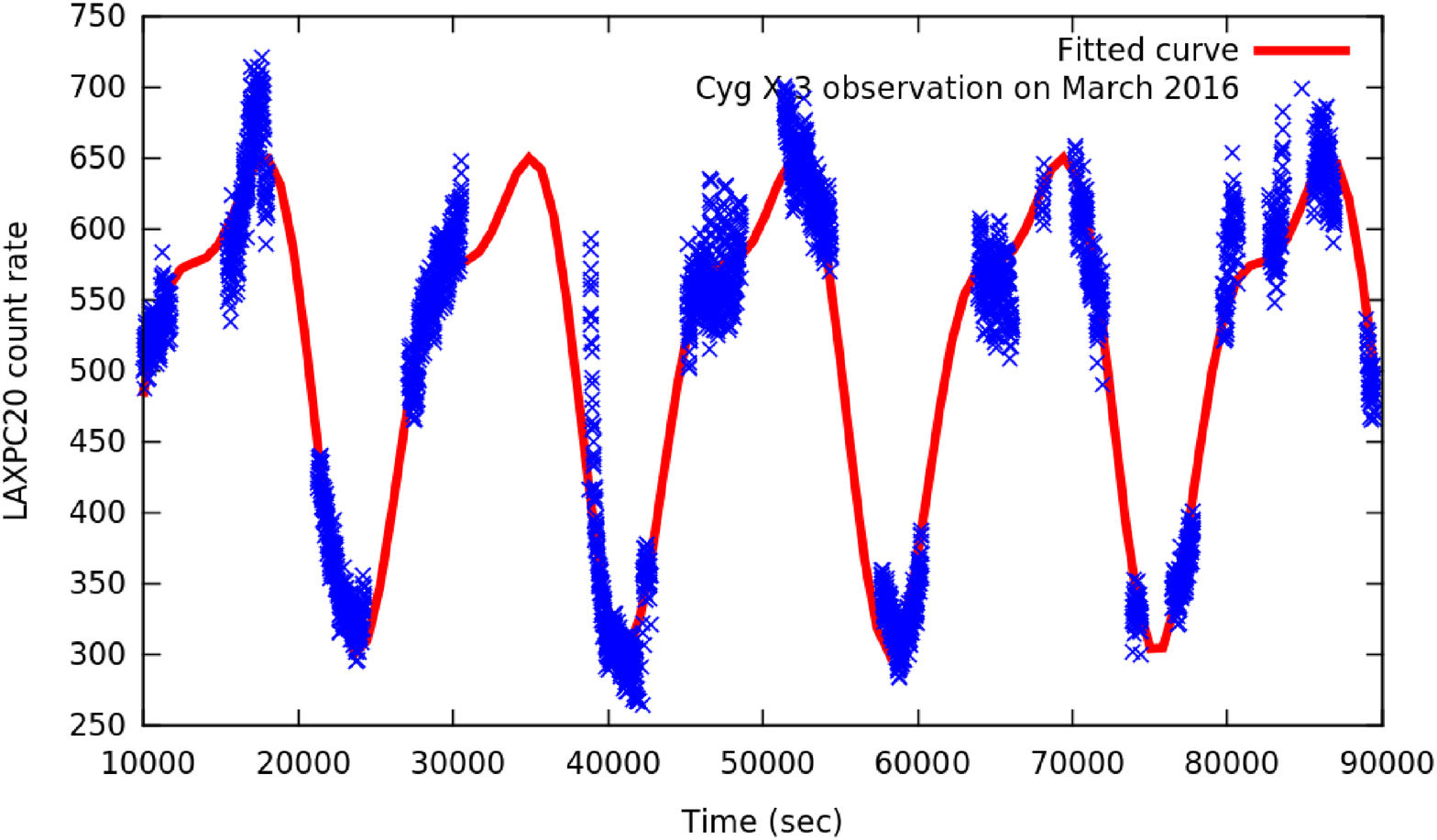}
\includegraphics[scale=0.245]{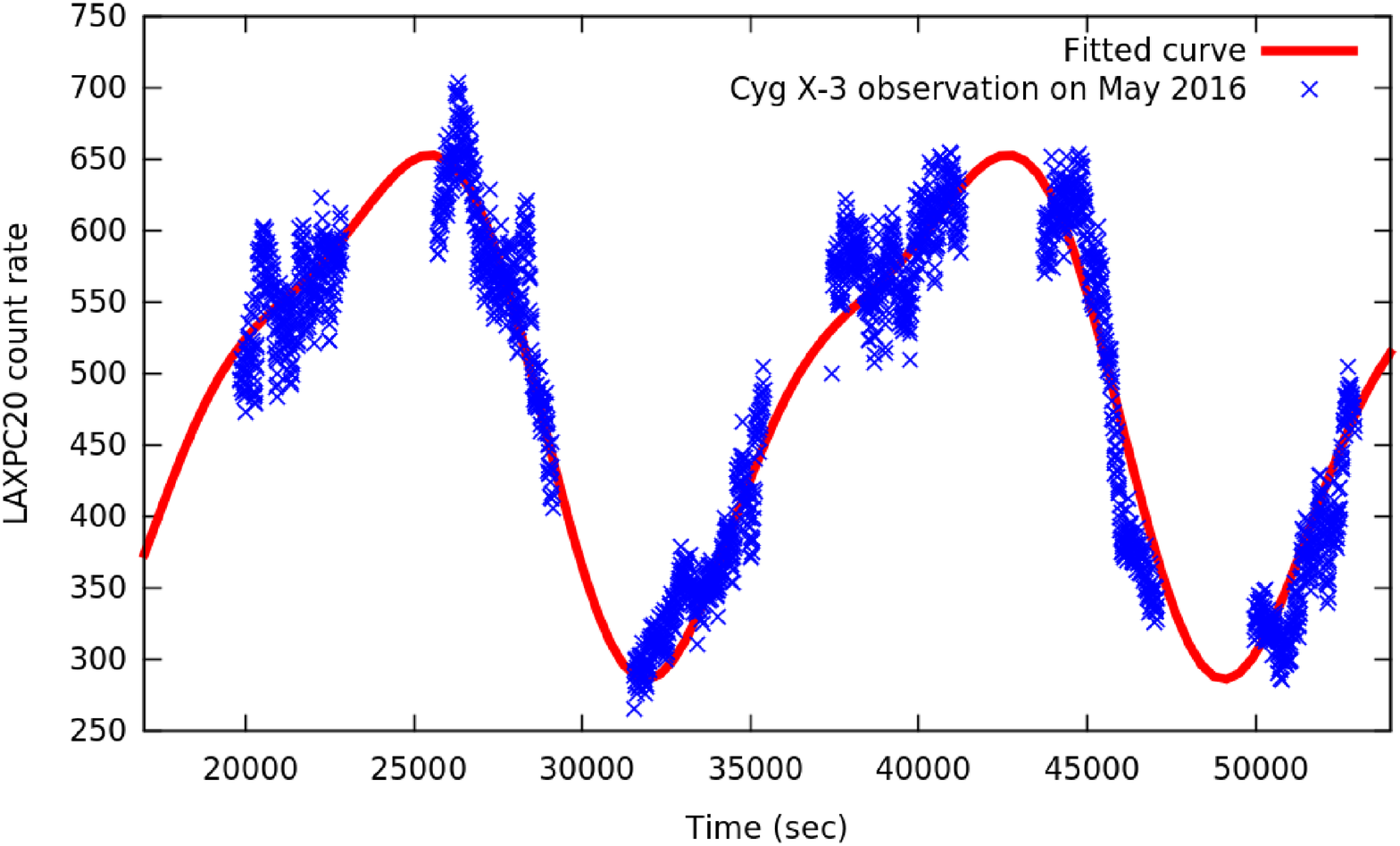}
\caption{Upper figure in the top left panel show background-subtracted LAXPC20 count rate (in blue squares) observed over many epochs covering one year of \asat{} observations. The variations can be modelled (in red) well with two sinusoids and its harmonics: one with binary orbital period of $\sim$4.8 hrs and another with precessional orbital period of $\sim$35 days. Central left shows the reduced $\chi^2$ ($\chi^2$/dof) of the sinusoidal fitting of the slowly varying component as a function of days. Two minima at $\sim$35.8 days and $\sim$38 days are obtained from the fitting. For clarity, the fitted sinusoidal model and its harmonic (in red) are zoomed along with background-subtracted 3-80 keV {\tt LAXPC20} lightcurve of Cyg X-3 (in blue) from consecutive orbits of several tens of ksec and shown in the top right and bottom panels during Nov 2015, Mar 2016 and May 2016 observations respectively.}
\label{fit}
\end{figure*}

\begin{table}
 \centering
 \caption{Orbital period measurements of Cyg X-3 using different epochs of LAXPC observations }
\begin{center}
\scalebox{0.80}{%
\begin{tabular}{ccc}
\hline 
LAXPC & average orbital & Precessional    \\
units & period (sec) & Period (days)  \\
\hline 
LAXPC10 & 17253.55 $\pm$ 0.18 & 35.7 $\pm$ 1.4 \\
LAXPC20 & 17253.49 $\pm$ 0.21 & 35.7 $\pm$ 1.3 \\
LAXPC30 & 17253.68 $\pm$ 0.11 & 36.0 $\pm$ 1.5 \\
\hline
All units combined & 17253.56 $\pm$ 0.19 & 35.8 $\pm$ 1.4 \\
\hline
zero phase reference & 57318.6545338 & $\pm$ 0.0000578 (MJD) \\
\hline 
\end{tabular}}
\end{center}
\label{period}
\end{table}

\begin{table*}
 \centering
 \caption{Details of QPO properties observed from Cyg X-3 during March 2016 observation }
\begin{center}
\scalebox{0.85}{%
\begin{tabular}{cccccccc}
\hline 
ID & Orbit number & Date &  Effective & background-subtracted & QPO frequency & QPO fractional & QPO detection  \\
no. & & (dd-mm-yyyy) &  exposure (sec) & source count rate & (mHz) & rms (\%) (3.0-80.0 keV) & significance ($\sigma$)\\
\hline 
Q1 & 2379 & 06-03-2016 & 2523.0 & 1943 $\pm$ 34 & 17.9 $\pm$ 0.5 & 0.97 $\pm$ 0.05 & $\sim$3.7 \\
Q2 & 2381 & 06-03-2016 & 3321.0 & 1648 $\pm$ 25 & 23.2 $\pm$ 0.9  & 1.47 $\pm$ 0.06 & $\sim$3.5 \\
Q3 & 2386B & 07-03-2016 & 3380.0 & 1715 $\pm$ 28 & 21.3 $\pm$ 0.4  & 1.23 $\pm$ 0.05 & $\sim$4.2 \\
NQ1 & 2380 & 06-03-2016 & 2953.0 & 962 $\pm$ 22 & --  & $\le$ 0.35$^a$ & -- \\
NQ2 & 2386A & 07-03-2016 & 2831.0 & 848 $\pm$ 19 &  -- & $\le$ 0.43$^a$ & -- \\
NQ3 & 2387 & 07-03-2016 & 2545.0 & 877 $\pm$ 19 &  -- & $\le$ 0.39$^a$ & -- \\
\hline
\end{tabular}}
\vspace{2ex}

\raggedright{\hspace{1.5cm}$^a$ the 3$\sigma$ upper limit of the fractional rms (per cents) measured from fitted PDS continuum (1$-$200 mHz) when}

\raggedright{\hspace{1.8cm}QPO is detected}
\end{center}
\label{qpo}
\end{table*}

\begin{figure}
\includegraphics[scale=0.33,angle=-90]{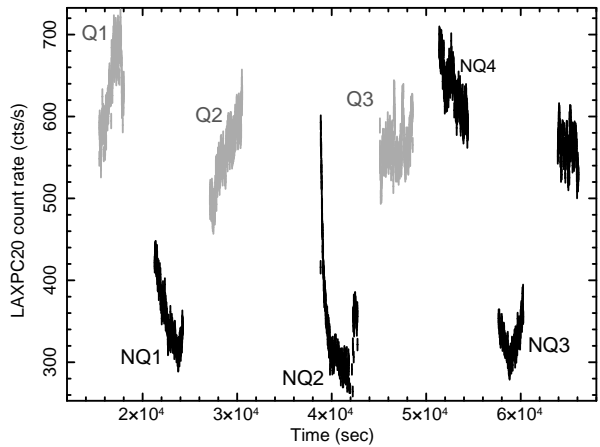}
\caption{LAXPC20 lightcurve of Cyg X-3 in the energy range 3.0-80.0 keV covering many \asat{} orbits continuously. The $\sim$20 mHz QPOs are detected during the rising phase of binary orbits (shown in grey) with at least 3$\sigma$ significances in three observations marked with `Q1', `Q2', and `Q3'. For clarity, observations where no QPOs are observed up to 2$\sigma$ significance and analysed here are marked as `NQ1', `NQ2', `NQ3' and `NQ4'. }
\label{mark}
\end{figure}

\section{Observations and analysis}

As a part of the payload verification (PV) phase calibration and guaranteed time (GT) observations, Cyg X-3 has been observed 9 times by \asat{}/LAXPC covering many orbits (between 24 October 2015 and 20 November 2016). Details of LAXPC observations during each epoch are shown in Table \ref{obs}. Spectral states provided in Table \ref{obs} are determined by qualitatively comparing the hardness intensity diagram (HID) with Figure 3 from \citet{ko10} and by comparing the unfolded spectral shape (in the unit of keV$^2$ (photons s$^{-1}$ cm$^{-2}$ keV$^{-1}$)) with Figure 4 from \citet{ko10}. Details of spectral states and corresponding broadband energy spectral analysis will be provided in a subsequent paper. We check each orbit data for possible detection of QPOs and observations where QPOs are detected with at least 99\% confidence are quoted in Table \ref{obs} with the number of detections. Observations on 6 March, 2016 have QPO detection significance higher than 3$\sigma$ and QPO fractional rms are highest among all detections. For this reason, we focus on the timing and spectral analysis of March data in this work. Remaining QPO detections have significance between 2.6-3$\sigma$. Using Equation 1, we calculate binary phase of the lightcurve intervals when QPOs are observed and provided them in the Table \ref{obs}. We also note that the binary phase corresponds to the maximum count rate is 0.64 $\pm$ 0.01. Therefore, all 7 QPOs are detected during the rising phase of the binary period during the flaring hard X-ray (FHXR) state. 
  
 All observations are taken in Event Analysis (EA) mode with the absolute time resolution of 10 $\mu$s in the energy range 3.0-80.0 keV. To check the source behavior, we also consider observations on 27 November 2015 and 21 May 2016 which are two epochs closest to March 2016 observations. During these three epochs of \asat{} observations, we use 15-50 keV \swift{}/BAT lightcurve to show hard flux behavior and the HID. Left panel of Figure \ref{hid} shows the one-day averaged BAT lightcurve where three epochs of observations are shown by vertical grey lines. It may be noted that the hard X-ray count rate During November 2015 was $\sim$60\% of that during March and May 2016. This indicates spectral hardening of the source occurred during March-May 2016 observation compared to November 2015 observation. This can be confirmed by the HID shown in the right panel of Figure \ref{hid}. The hardness is defined as the ratio of count rate in the energy range 10-30 keV and 3-6 keV. Such definitions of hard and soft bands have been used previously on few occasions \citep{ko11, du10, fe04} and ensures mutually exclusive hard and soft bands in black hole X-ray binaries. The HID of November 2015 and March/May 2016 form parallel tracks in the HID and Nov 2015 HID is significantly softer (nearly by a factor of 2) than the HID observed on March and May 2016. Comparing with \citet{ko10}, we find that March and May 2016 observations belong to the flaring hard X-ray state (FHXR) while the November 2015 observations belong to the flaring intermediate state (FIM). Such trends of parallel tracks have also been observed with \xte{} (see Figure 1 of \citet{ko10}). During harder state and at the rising X-ray flux, 20 mHz QPOs are observed, the HID position of which is shown by red symbols in the Figure \ref{hid}. 

Based on good time intervals, when lightcurves from several consecutive orbits over one year are combined covering many epochs as observed from Table \ref{obs}, a strongly periodic modulation is observed. A careful observation of the lightcurve from different epochs show both sinusoidal binary motion and its harmonics. Not only that, a slower variation with the time-scale of the order of $\sim$ 5 weeks are also observed. With this motivation, we assume lightcurves from all epochs can be modelled with a function consisting of the addition of two sinusoids and their harmonics: one with faster variability of the order of binary orbital period and the another with the slower variability of the order of a month time-scale. The functional form of the model is :

\begin{multline}
f(t) = a_0 + a_1 \sin(\omega_b t) + a_2 \cos(\omega_b t) + a_3 \sin (2 \omega_b t)\\
 + a_4 \cos (2 \omega_b t) + a_5 \sin(\omega_p t) + a_6 \cos(\omega_p t) \\
 + a_7 \sin (2 \omega_p t) + a_8 \cos (2 \omega_p t) 
\end{multline}

where $\omega_b$ and $\omega_p$ represent the orbital and precessional angular velocities and a$_0$ to a$_8$ are coefficients. In stead of addition, we also fit with the function consists of the multiplication of two sinusoids and their harmonics assuming that the slower and the faster varying components are originated from the same physical process. The multiplicative function is not able to fit the observed profile. The reason is that the product function have a constant mean value when averaged over shorter period, which is contrary to observed increase in the mean value at least, during one observation (see top left panel of Figure \ref{fit}). This follows from the fact that if this signal is averaged over a period of shorter time-scale ($\sim$ 4.8 hours), then the slowly varying component (of the order of $\sim$ 5 weeks) will hardly change on that time-scale, while the $\sim$4.8 hours factor will average to zero. Thus only the constant term will contribute to the average. The proposed model is fitted to all 9 observations of Cyg X-3 shown in the top left panel of Figure \ref{fit}.

In order to determine the slowly varying component, $\omega_p$, we have taken the entire light curve spanning $\sim$400 days with gaps with a time bin of 60 sec that will have a few thousand points and the fit involves 9 coefficients in equation 1 and fixed $\omega_p$. The fit is done separately for each value of $\omega_p$ with the step size of 0.1 days between 32 and 42 days and for each step of $\omega_p$, $\omega_b$ is varied in small steps (0.01 sec) around its best-fit values so that we can get both simultaneously, by looking for minimum $\chi^2$. The resulting fit is plotted in the central left panel of Figure \ref{fit}, which shows $\chi^2$/dof as a function period. The minimum of $\chi^2$/dof obtained from variation of both periodicities is taken to be the best value giving the orbital and precession periodicities.

From the fitted parameters of the additive function given in equation 1, we obtain the combined (combining measurements from three LAXPC units) orbital periodicity to be 17253.56 $\pm$ 0.19 sec which is consistent with the measured orbital period of the system and a slower variation of 35.8 $\pm$ 1.4 days which is consistent with the periodicity of $\sim$34.1 days observed earlier from this source \citep{ho76, mo80} and it was explained using a free precession model of an elastic neutron star \citep{su85}. Table \ref{period} provides orbital and precessional periodicities obtained from three LAXPC units individually and combined. The zero phase reference (57318.6545338 $\pm$ 0.0000578 MJD) corresponds to the first minima of the binary orbital period. The model-fitted curve and LAXPC20 count rate covering all epochs (provided in Table \ref{obs}) spanning over nearly a year are shown in the top left panel of Figure \ref{fit}. In the central left panel of Figure \ref{fit}, the reduced $\chi^2$ of the fitting of slowly varying component show two apparent minima at $\sim$35.8 days and $\sim$38.2 days respectively. The reason for two periodicities is not clear however, the minima at $\sim$35.8 days is lower than the minima at $\sim$38.2 days, more uniform and consistent with previous measurements. Therefore, in the Table \ref{period} we quote measurements at $\sim$35.8 days from three LAXPC units. 
It may be noted that we cannot rule out the possibility that one of the slow periodic variability is due to the window function convolution with the other period. However, putting random values for the observed points or taking random combinations of 9 sets of observations do not give any significant periodicity in this interval. But in order to determine the long period reliably, we  need many more data set, so we can only drop the first few or the last set to check the effect of the window function. With just 9 data set, it is also difficult to rule out the presence of systematic variation in the long period. However, if the period is varying steadily then one would expect the $\chi^2$/dof dip to be broadened and the width of dips can give some measure of variation. Since within $\sim$395 days of observations, about 10-11 periods are covered, we would expect a width of about 10\%, which can include both dips in $\chi^2$/dof vs period plot shown in the central left panel of Figure \ref{fit}. The cadence of {\it Swift}/BAT lightcurve is not small enough to remove the effect of smaller period. Hence it is difficult to conclude systematic variations in the long-period oscillation from the BAT lightcurve.

For clarity, we show the zoomed version of the model along with {\tt LAXPC20} count rate during three epochs of observations covering consecutive orbits of few tens of kilo-seconds in top right and bottom panels of Figure \ref{fit}.

Out of several orbits, we present power density spectral analysis of six LAXPC observations of Cyg X-3 during March, 2016 as listed in Table \ref{qpo}. Each of them were obtained from a single orbit with effective exposure of $\sim 2-3$ ksec each. In three orbits, a quasi-periodic oscillation (QPO) is detected significantly in the power density spectra (PDS) which are denoted by `Q1', `Q2', `Q3' respectively while observations from three orbits where no such QPO-like features are observed are denoted by `NQ1', `NQ2' and `NQ3' in Figure \ref{mark} and Table \ref{qpo} respectively. 

\begin{figure*}
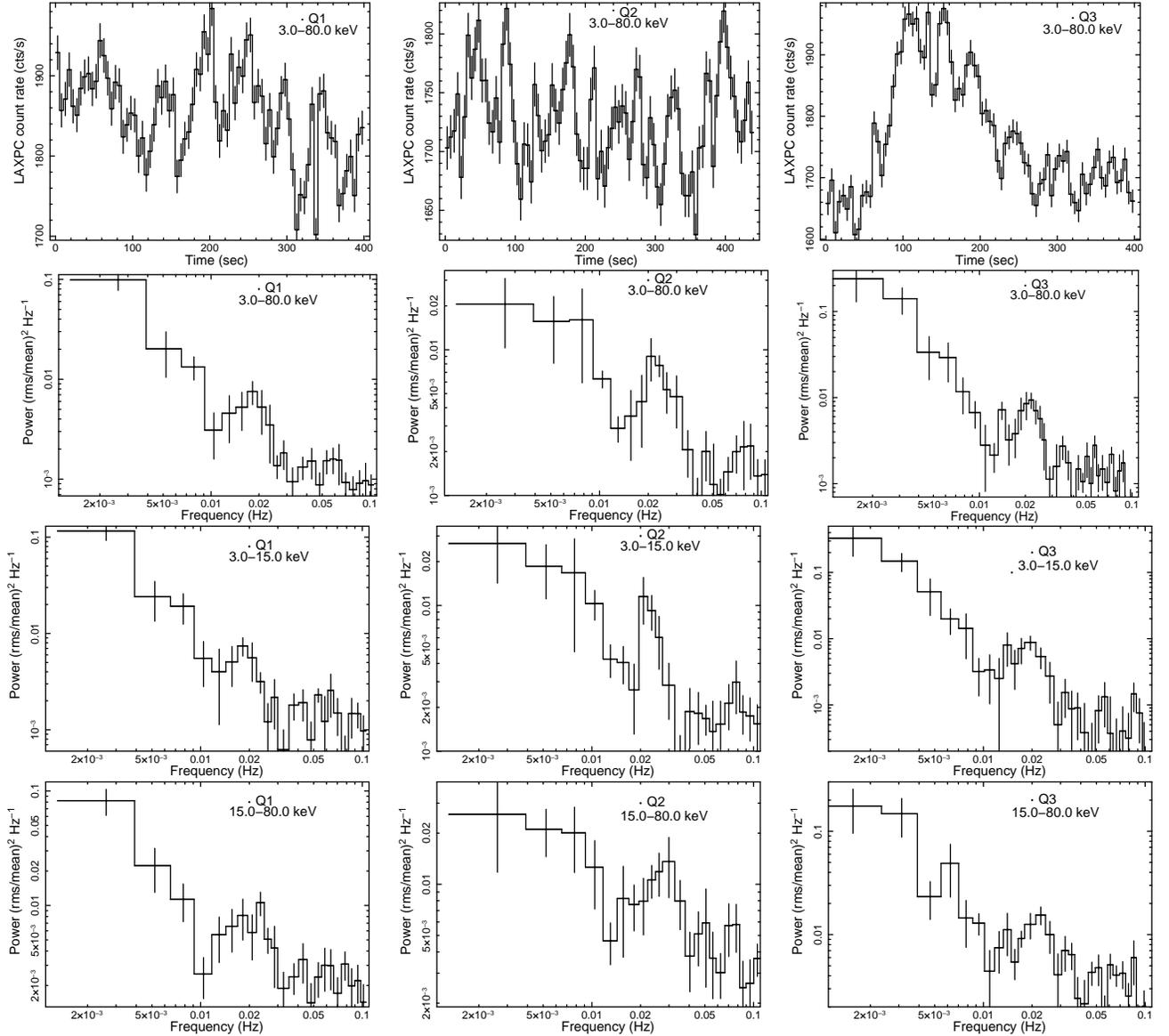

\centering \includegraphics[width=0.22\textwidth,angle=-90]{fig4a.ps}
\centering \includegraphics[width=0.22\textwidth,angle=-90]{fig4b.ps}
\centering \includegraphics[width=0.22\textwidth,angle=-90]{fig4c.ps}
\centering \includegraphics[width=0.21\textwidth,angle=-90]{fig4d.ps}
\centering \includegraphics[width=0.21\textwidth,angle=-90]{fig4e.ps}
\centering \includegraphics[width=0.21\textwidth,angle=-90]{fig4f.ps}
\centering \includegraphics[width=0.21\textwidth,angle=-90]{fig4g.ps}
\centering \includegraphics[width=0.21\textwidth,angle=-90]{fig4h.ps}
\centering \includegraphics[width=0.21\textwidth,angle=-90]{fig4i.ps}
\centering \includegraphics[width=0.21\textwidth,angle=-90]{fig4j.ps}
\centering \includegraphics[width=0.21\textwidth,angle=-90]{fig4k.ps}
\centering \includegraphics[width=0.21\textwidth,angle=-90]{fig4l.ps}
\caption{Top panels in three columns show $\sim$400 sec section of lightcurve from each of observations (Q1, Q2 and Q3) where $\sim$20 mHz QPOs are detected. For better visibility of $\sim$50 sec (1/20 mHz) oscillations, 400 sec segments are used. Lightcurves are binned to 5 sec. Second rows of panels show 3.0-80.0 keV Poisson-noise subtracted power density spectra from Q1, Q2 and Q3  (left to right). Third and fourth rows of panels show PDS in the the energy range 3.0-15.0 keV and 15.0-80.0 keV respectively. All PDS are Poisson-noise subtracted and derived from observations combining all three LAXPC units. In three energy ranges: 3.0-15.0 keV, 15.0-80.0 keV and 3.0-80.0 keV, $\sim$20 mHz QPOs are observed from all PDS with variable strength.}
\label{power}
\end{figure*}

\section{Timing analysis and results}

A close inspection of Figure \ref{mark} and Table \ref{obs} reveal that QPOs are detected mostly in the observations that were obtained during the rising flux of the orbital phases. A 400 sec section of lightcurve of three observations $-$ 'Q1', `Q2' and `Q3' is shown in the top panels of Figure \ref{power} respectively. Lightcurves from all three LAXPC units are combined and background subtracted. A bin size of 5 sec is used. A quasi-periodicity in low-amplitude flares of the order of $\sim$40-60 sec can be observed from all three lightcurves in Figure \ref{power}. Top right panel shows $\sim$50 sec flares on the top of a large flare of the order of $\sim$300 sec.  

\begin{figure*}
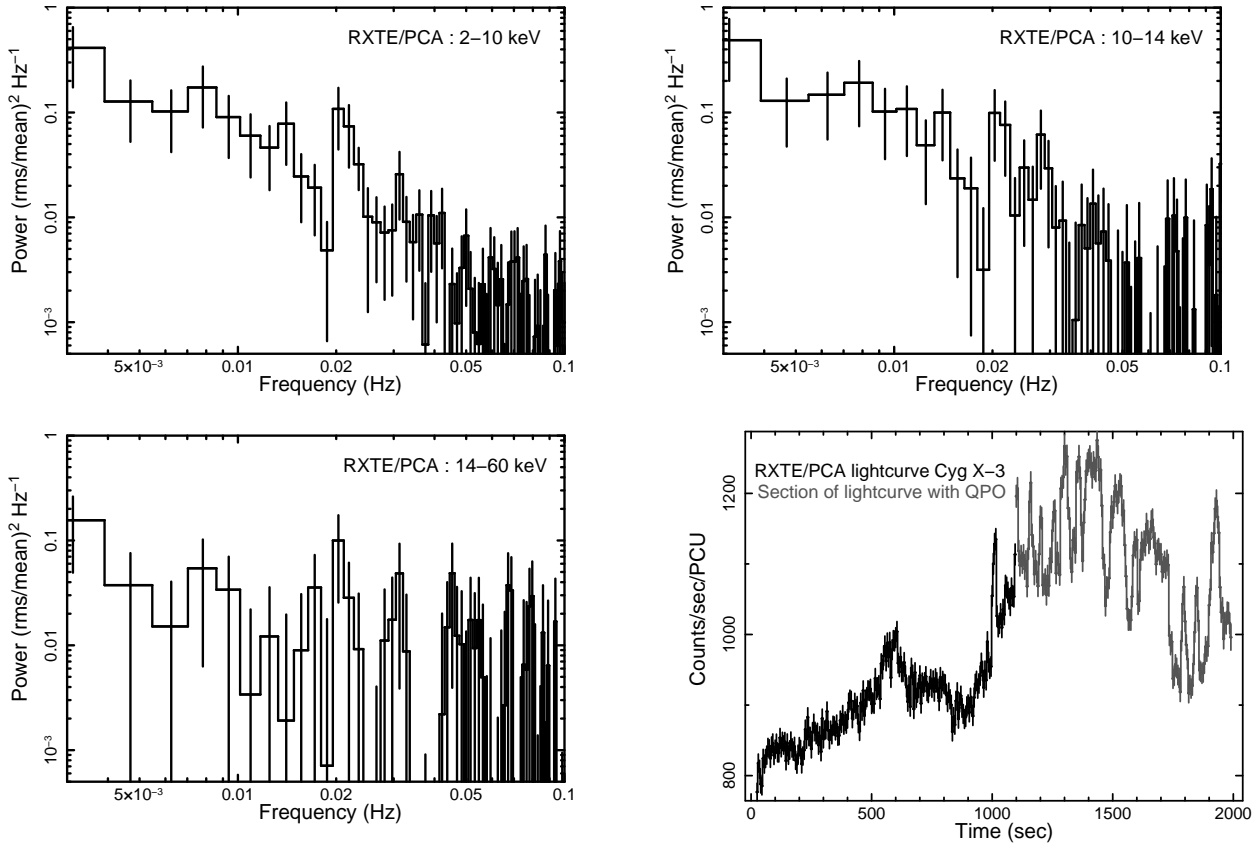

\begin{tabular}{cc}
\includegraphics[scale=0.31,angle=-90]{fig5a.ps} &
\includegraphics[scale=0.31,angle=-90]{fig5b.ps} \\
\includegraphics[scale=0.31,angle=-90]{fig5c.ps} &
\includegraphics[scale=0.31,angle=-90]{fig5d.ps} \\
\end{tabular}
\caption{\xte{}/PCA observation of Cyg X-3 on 09 August, 2009: top left panel shows 2-10 keV Poisson-noise subtracted power density spectra (PDS) where $\sim$21.8 mHz quasi-periodic oscillation (QPO) is observed at the significance of $>$ 3$\sigma$. The PDS in 10-14 keV and 14-60 keV energy ranges are shown in the top right and bottom left panels where no QPO at $\sim$21.8 mHz is detected up to 3$\sigma$. 2-60 keV lightcurve is shown in the bottom right panel where the section of lightcurve that shows $\sim$21.8 mHz QPO is marked in grey. The binary phase of the \xte{}/PCA lightcurve when the QPO is observed is 0.231-0.335 as calculated using equation 1.}
\label{pca}
\end{figure*}

\begin{figure*}
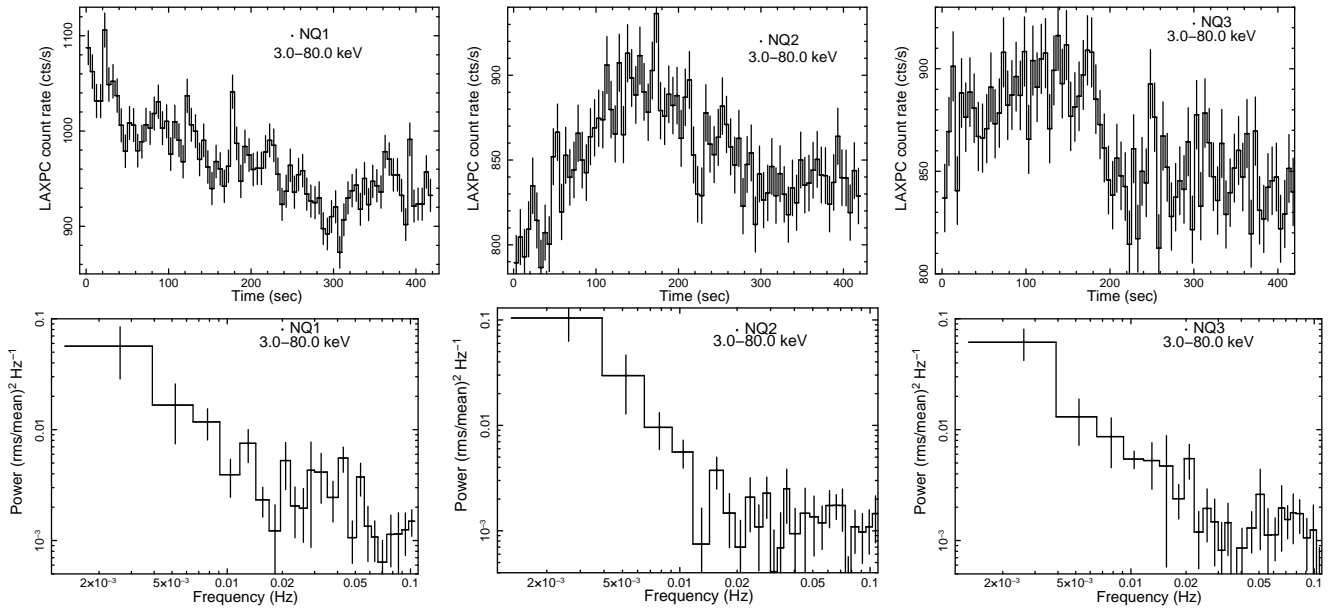

\centering \includegraphics[width=0.22\textwidth,angle=-90]{fig6a.ps}
\centering \includegraphics[width=0.22\textwidth,angle=-90]{fig6b.ps}
\centering \includegraphics[width=0.22\textwidth,angle=-90]{fig6c.ps}
\centering \includegraphics[width=0.22\textwidth,angle=-90]{fig6d.ps}
\centering \includegraphics[width=0.22\textwidth,angle=-90]{fig6e.ps}
\centering \includegraphics[width=0.22\textwidth,angle=-90]{fig6f.ps}
\caption{Top panels in three columns show $\sim$400 sec section of lightcurve from each of observations (NQ1, NQ2 and NQ3) where no QPOs are detected between 1 mHz and 100 mHz. For comparison with Figure \ref{power}, 400 sec segments are used. Lightcurves are binned to 5 sec. All lightcurves are background-subtracted and combined using three LAXPC units. Second rows of panels show 3.0-80.0 keV Poisson-noise subtracted power density spectra from NQ1, NQ2 and NQ3 observations (left to right) where no obvious features are present. }
\label{nq}
\end{figure*}

\subsection{LAXPC rising-phase PDS analysis}

Power density spectra (PDS) are derived from `Q1', `Q2' and `Q3' lightcurves in the energy range 3.0-80.0 keV, 3.0-15.0 keV and 15.0-80.0 keV and shown in the second, third and fourth rows of panels in the Figure \ref{power}. Combining lightcurves from three LAXPC units in three energy bands, PDS are derived. All PDS are rms-normalized and deadtime-corrected Poisson-noise subtracted. Details of deadtime-corrected Poisson noise estimations are discussed in \citet{ya16a}. PDS are plotted in the frequency range 1 mHz to 100 mHz using suitable geometric re-binning for the clarity of any feature. In Figure \ref{power}, for all three PDS in 3.0-15.0 keV (all panels in third row) and 15.0-80.0 keV (bottom panels) energy bands, a QPO-like feature is observed at $\sim$20 mHz. To determine strength and significances of such features, the PDS continuum are fitted with a broken powerlaw while QPO-like feature is fitted with Lorentzian. QPO frequencies as obtained from fitted Lorentzian centres are quoted in Table \ref{qpo}.

The significances of such QPO-like features are computed by dividing the area under the Lorentzian at QPO position with the 1$\sigma$ negative error of model-estimated area. Detection significances of QPOs in `Q1', `Q2' and `Q3' are quoted in Table \ref{qpo}. We calculate the confidence level of detected QPOs following the recipe for testing the significance of peaks in the periodogram of red noise data provided by \citet{va05}. If the red noise can be fitted by a powerlaw-like continuum, then low significance peaks can be rejected accurately using this recipe. We find that the peak X-ray power in `Q1', `Q2' and `Q3' PDS at the QPO frequency is higher than the power predicted at least at 99.0\% confidence level. Using the fitted Lorentzian parameter, fractional rms is estimated and then they are corrected for the background by multiplying by the factor R/S; where R is the raw, observed count rate and S is the background-subtracted source count rate. 3.0-80.0 keV fractional rms of `Q1', `Q2' and `Q3' are quoted in Table \ref{qpo}.

\begin{figure*}
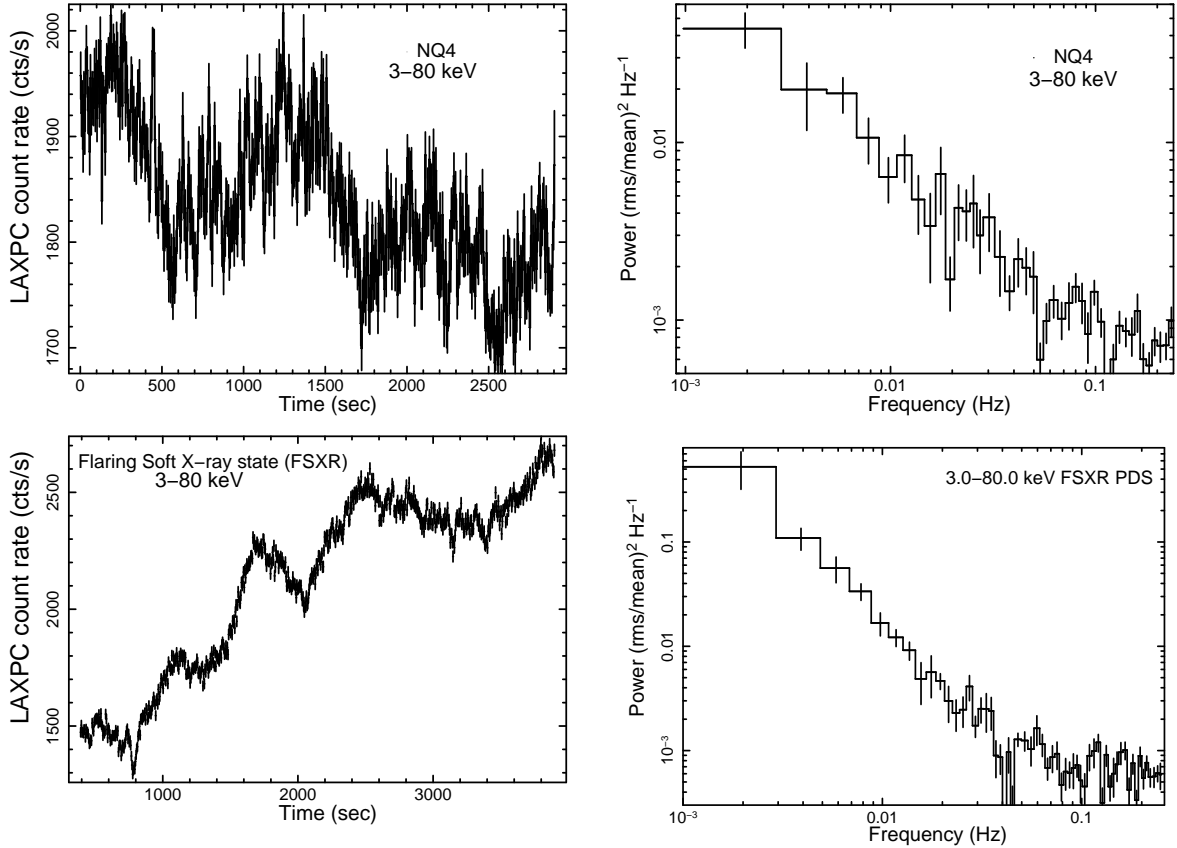

\begin{tabular}{cc}
\includegraphics[scale=0.31,angle=-90]{fig7a.ps} &
\includegraphics[scale=0.31,angle=-90]{fig7b.ps} \\
\includegraphics[scale=0.31,angle=-90]{fig7c.ps} &
\includegraphics[scale=0.30,angle=-90]{fig7d.ps} \\
\end{tabular}
\caption{Top left and right panels show the 5.0 sec binned lightcurve of Cyg X-3 during the decaying phase of the binary orbital motion (shown by `NQ4' in Figure \ref{mark}) and the corresponding power density spectrum respectively. Despite of high count rate and sufficient exposure, no QPO-like feature is visible in the poisson-noise subtracted and background-corrected PDS. This implies that the detection of QPO is exclusive to the rising phase of binary orbital motion. Bottom left and right panels show the 5.0 sec binned lightcurve of Cyg X-3 during the rising binary phase of the flaring soft X-ray state (FSXR) and the corresponding power density spectrum respectively. Despite of the rising phase in FSXR state and sufficient exposure, No QPO-like feature is visible in the poisson-noise subtracted and background-corrected PDS. This implies the 20 mHz QPO may be state dependent.}
\label{special}
\end{figure*}

\subsubsection{\xte{}/PCA PDS analysis}

As pointed by \citet{ko11}, during flaring hard X-ray state, a $\sim$22 mHz QPO from Cyg X-3 is detected at the significance level higher than 99 per cent with \xte{}/PCA observation on 09 August, 2009. We reanalyze the same observation and extract Poisson-noise subtracted PDS in three different energy bands: 2-10 keV, 10-14 keV and 14-60 keV. The choice of bands are restricted by the PCA default channel binning. Top left panel of Figure \ref{pca} shows the PDS in 2-10 keV where a strong QPO-like feature is observed at $\sim$22 mHz. We fit the PDS with a combination of powerlaw and Lorentzian and compute the significance of the QPO using the method described above. The fit returns the QPO at 22.8 $\pm$ 0.5 mHz with the QPO significance of 3.3$\sigma$. Interestingly, no Lorentzian are required to improve the fit significantly at the position of the QPO frequency in the PDS extracted at 10-14 keV (top right panel of Figure \ref{pca}) and 14-60 keV (bottom left panel of Figure \ref{pca}). QPO significance at $\sim$22 mHz in 10-14 keV and 14-60 keV are $<$ 3$\sigma$ and $<$ 2$\sigma$ respectively. This is in sharp contrast with \asat{}/LAXPC observation of $\sim$21 mHz at higher energy. In 15-80 keV energy range, QPOs are detected at the significance of $>$ 3$\sigma$. The only reason of the detection of QPO at high energy with LAXPC and not with PCA is the efficiency of LAXPC which is higher by few factors than PCA above 20 keV (at 20 keV, the effective area of \xte{}/PCA is $\sim$1500 cm$^2$ \citep{ja06} while the effective area of \asat{}/LAXPC is $\sim$6000 cm$^2$ \citep{an17}). The section of \xte{}/PCA lightcurve that show mHz QPO is plotted in gray in the bottom right panel of Figure \ref{pca}. The binary phase of the time interval when \xte{}/PCA QPO is observed is calculated in the range of 0.231-0.335 using Equation 1. This phase interval also belongs to the rising phase of the binary motion and the range partially overlaps with a couple of QPO detections with LAXPC (see Table \ref{obs}).            

\begin{figure*}
\centering \includegraphics[width=0.25\textwidth,angle=-90]{fig8a.ps}
\centering \includegraphics[width=0.25\textwidth,angle=-90]{fig8b.ps}
\centering \includegraphics[scale=0.26]{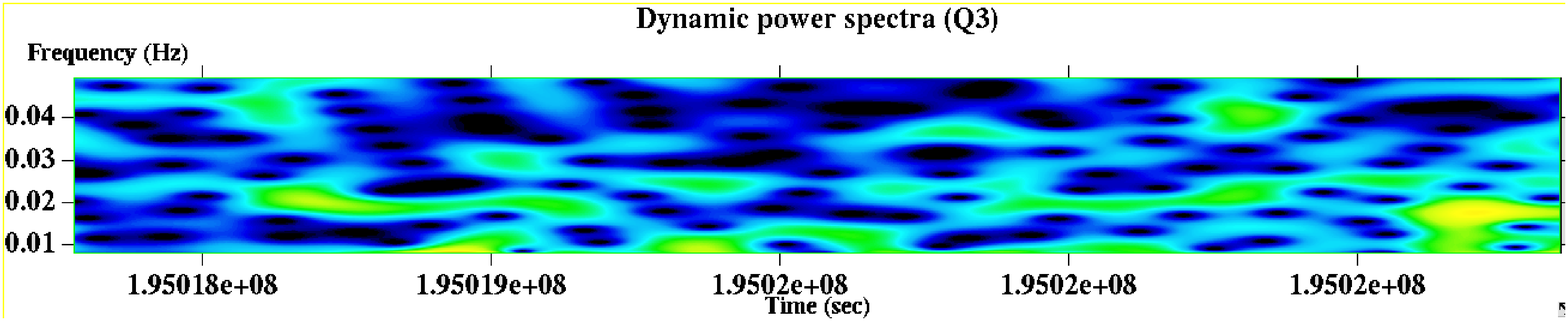}
\centering \includegraphics[scale=0.26]{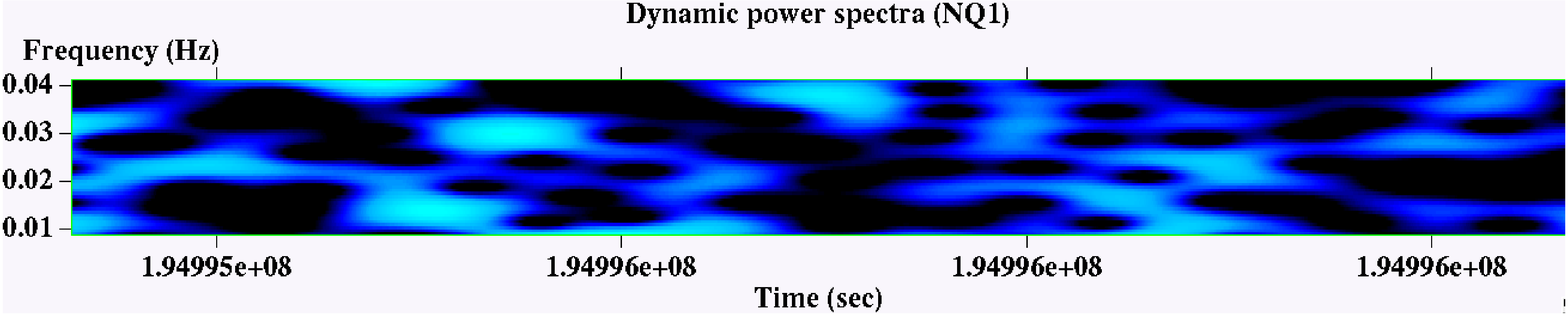}
\centering \includegraphics[width=1.2\columnwidth,angle=-0]{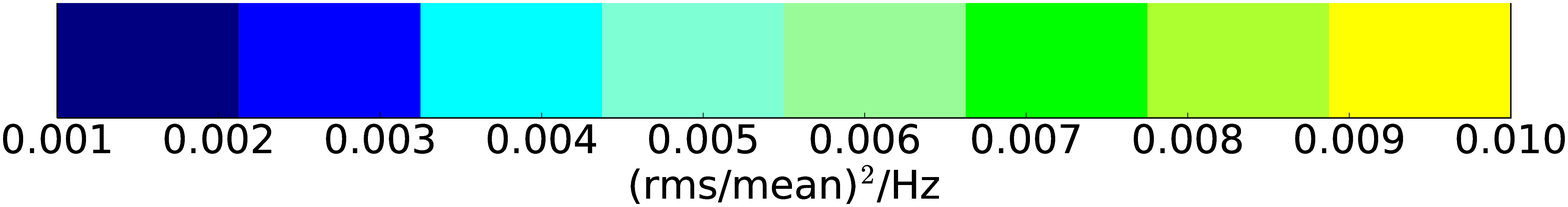}
\caption{Top left panel shows two Poisson-noise subtracted power density spectra (PDS) plotted together when QPO is detected (black) and no QPO is detected (red). Top right panel shows PDS during Q3 observation in the energy range 3.0-80.0 keV fitted with a combination of a broken powerlaw and two Lorentzians along with its residual. Middle panel shows dynamic power spectra (DPS) from the lightcurve when QPO is detected (Q3) and bottom panel shows DPS from the lightcurve when no QPO is detected. Clearly a semi-persistent QPO is detected in the middle panel at $\sim$20 mHz which is absent from the bottom panel. }
\label{dps}
\end{figure*}

\subsection{LAXPC decay phase PDS analysis}

A $\sim$400 sec section of lightcurve of three observations $-$ `NQ1', `NQ2' and `NQ3' are shown in the top panels of Figure \ref{nq} respectively. Similar procedures are used to extract lightcurves. None of the lightcurves show low-amplitude, quasi-periodic $\sim$40-60 sec flare-like structures although large flares of the order of $\sim$300 sec is observed in `NQ2' lightcurve, similar to what is observed from `Q3' lightcurve.
Following previous methods, we derive rms-normalized, deadtime-corrected Poisson-noise subtracted PDS from `NQ1', `NQ2' and `NQ3' lightcurves in 3.0-80.0 keV energy range. Bottom panels in Figure \ref{nq} show PDS from `NQ1', `NQ2' and `NQ3' respectively in the frequency range from 1 mHz to 100 mHz. Although strong noise component is observed below $\sim$15 mHz which rises at lower frequencies, no QPO-like features are observed from all PDS at $\sim$20 mHz. 
In order to show that the low source count rate during `NQ1', `NQ2' and `NQ3' (from top panels of Figure \ref{nq}) and non-detection of QPOs are not correlated, we select a high count rate observation (comparable to that when QPO is detected) during the decay phase of the binary motion and marked by `NQ4' in the Figure \ref{mark}. The 5 sec binned lightcurve with the exposure of $\sim$3 ks is shown in the top left panel of Figure \ref{special} and the corresponding power density spectrum is shown in the top right panel. No QPO is detected in the PDS implies that QPO detection does not depend on source count rate in Cyg X-3 and QPOs are observed mostly during the rising phase of the binary orbital motion. In order to show whether mHz QPOs are spectral state dependent, we consider the rising phase lightcurve of the flaring soft X-ray state (FSXR) which is shown in the bottom left panel of Figure \ref{special} and its corresponding PDS in the bottom right panel. The PDS shows no signature of mHz oscillations which probably imply that $\sim$20 mHz QPO is more common in the FHXR state.    

To compare PDS with and without QPOs, we plotted them together in the top left panel of Figure \ref{dps}. It clearly shows no QPO during the lightcurve of the decay orbital phase while a strong QPO-like feature is present in the lightcurve during the rising orbital phase. A PDS during `Q3' fitted with a combination of broken powerlaw and Lorentzians is shown in the top right panel of Figure \ref{dps} where different model components and residual of the fit are also shown. 3.0-80.0 keV dynamic power spectra (DPS) of lightcurve with (`Q3') and without (`NQ1') QPO are shown in the bottom panels of Figure \ref{dps}. DPS are extracted using {\tt fgabor} tool in {\tt FTools v 6.18} which performs a Gabor transformation on a lightcurve and return the normalized power as a function of time and frequency. In the frequency range 10-40 mHz, the DPS of `Q3' in Figure \ref{dps} shows strong excess around 20 mHz ((rms/mean)$^2$ Hz$^{-1}$ $>$ 0.0085) although it is quasi-continuous and the power varies with time. Such strong excess is absent from the DPS of `NQ3' where (rms/mean)$^2$ Hz$^{-1}$ $<$ 0.0033 with 3$\sigma$ significance.

\begin{figure}
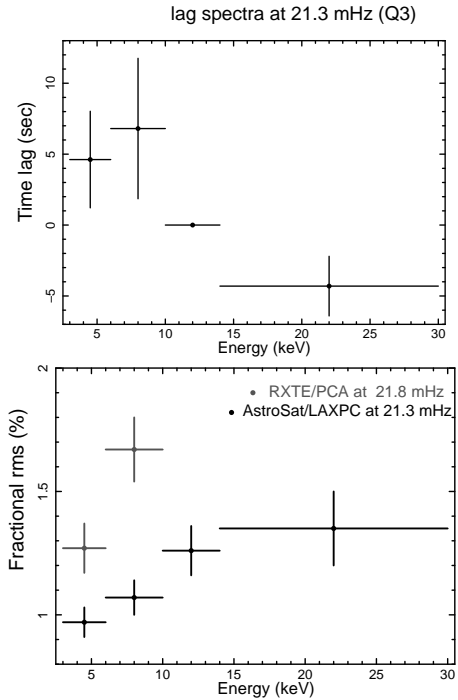

\centering \includegraphics[width=0.26\textwidth,angle=-90]{fig9a.ps}
\centering \includegraphics[width=0.25\textwidth,angle=-90]{fig9b.ps}
\caption{Top panel shows time-lag spectra as a function of photon energy as observed from \asat{}/LAXPC at the QPO frequency of 21.3 mHz (Q3). While calculating time delay in different energy bands, 10.0-14.0 keV is considered as the reference band. A soft-lag behaviour (soft photons lag hard photons) is observed. Such study with \xte{}/PCA is not possible since QPO is not detected above 10 keV. Bottom panel shows background-corrected fractional rms (\%) of the mHz QPO from \asat{}/LAXPC (shown in black) and \xte{}/PCA (shown in grey) as a function of photon energy. Since QPO is not detected above 10 keV with \xte{}/PCA, the rms spectral study with PCA is restricted to 10 keV.}
\label{lag}
\end{figure}

\subsection{Time-lag and rms spectra}

Event mode data from LAXPC distributed over 1024 channels allows to compute time-delay between photons at different energies. Time-lag are calculated at $\sim$21.3 Hz QPO frequency during `Q3' observation, in which the QPO detection significance is highest. 10-14 keV energy band is selected as the reference band since the QPO fractional rms at this energy range is moderately high. Details of time-lag calculation and its error estimations are provided in \citet{no99}, Time-lag spectra is shown in the top panel of Figure \ref{lag}. At higher energy ($>$ 7 keV), a significant soft lag, where soft X-ray photons lag hard X-ray photons, is observed. Background-corrected fractional rms of the QPO observed from both \xte{}/PCA and \asat{}/LAXPC are calculated at different energy bands and shown using grey and black symbols respectively in the bottom panel of Figure \ref{lag}. Fractional rms is found to increase with photon energy from both instruments. With \asat{}/LAXPC observations, energy-dependent lag and rms calculations are restricted to 30 keV due to poor signal-to-noise ratio at higher energies. 

\section{Spectral analysis and results}

To find out the change in spectral properties while the source moves from the rising orbital phase to the decay orbital phase and to connect such changes with the appearance and disappearance of QPOs at the rising and decay orbital phase, we perform average energy spectral analysis in the energy range 4.0-60.0 keV of `Q1' observation which resides at the rising phase and `NQ1' which resides at decay phase, immediately after `Q1' (see Figure \ref{mark}). The response function for each of the LAXPC units is computed using {\tt GEANT4} simulations \citep{an17,ya16b} and the background for each of the unit is modelled using blank sky observation at different position of the satellite in its orbit \citep{an17}. A 4\% uncertainty on the model background count rate is added and 1.5\% model systematic error is introduced. The spectral fitting is performed using the {\tt XSPEC v 12.9.0n} fitting package.  

\begin{figure}
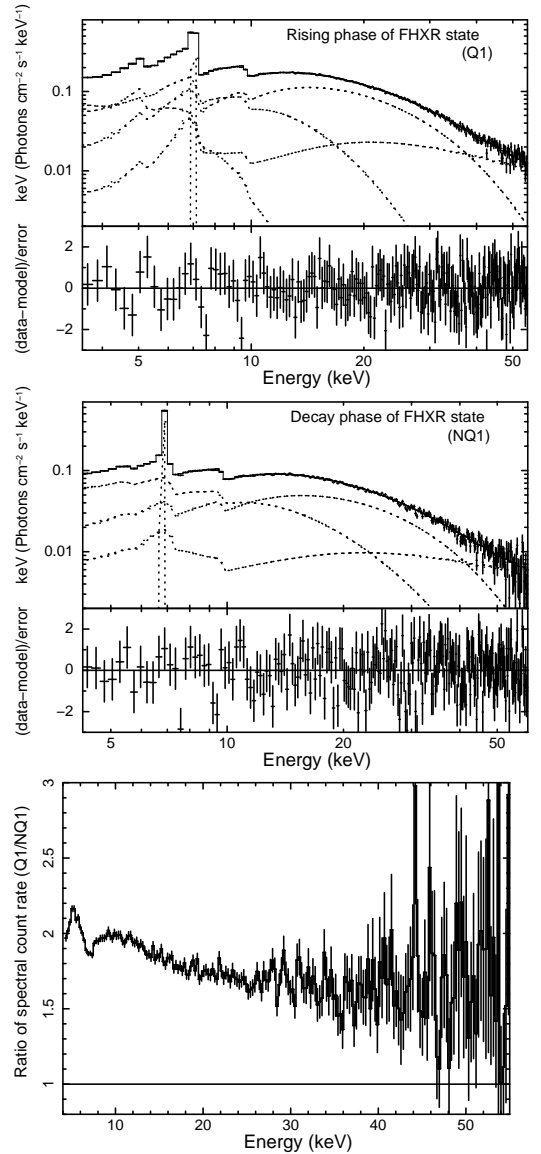

\centering \includegraphics[width=0.28\textwidth,angle=-90]{fig10a.ps}
\centering \includegraphics[width=0.28\textwidth,angle=-90]{fig10b.ps}
\centering \includegraphics[width=0.28\textwidth,angle=-90]{fig10c.ps}
\caption{The top panel shows the energy spectra of Cyg X-3 along with its residual obtained during the rising phase when fitted with a model consisting of a disk blackbody emission, thermal Comptonization, thermal Bremsstrahlung emission, disk reflection, a narrow emission line and edges. Middle panel shows fitted energy spectra of Cyg X-3 along with the residual during decay phase using the same model. Bottom panel shows the ratio of background-subtracted rising phase and decay phase spectral count rate as a function of photon energy. Up to 50 keV, the ratio is found to be significantly higher than 1. Such behavior of the ratio spectrum indicate a change in spectral shape between the rising and decay phase spectra.}
\label{spec}
\end{figure}

It has already been noticed several times that an absorbed thermal Comptonization is the predominant component in the energy spectra \citep{ko13,vi03}. To fit energy spectra during `Q1' and `NQ1', we use the same best-fit model as described in \citet{ko13} except the {\tt BELM} thermal/non-thermal hybrid Comptonization model. \citet{ko13} used highly sensitive, broadband spectra in the energy range 1-100 keV from multiple instruments. Therefore, the complex, self-consistent, hybrid model like {\tt BELM} which in addition to the thermal/non-thermal Comptonization, self-consistently consider disk spectrum as the origin of soft seed photons as well as compute reflection fraction from the disk. It is suitable for describing broadband, high quality spectra. However, our spectral analysis is restricted to 4-60 keV and the spectra are not sensitive enough to probe hybrid corona. Therefore, we find that in stead of the complex model like {\tt BELM}, a simpler model consisting of a disk blackbody ({\tt diskbb} in {\tt xspec}), a thermal Comptonization ({\tt nthcomp} in {\tt xspec}) and a disk reflection model {\tt reflionx} \citep{ro05} can describe the LAXPC spectra well in 4-60 keV. We replace the absorption model {\tt phabs} with {\tt tbabs}. With the addition of a narrow Gaussian emission line at 6.97 keV and two edges at $\sim$5.8 keV and $\sim$9.6 keV, the spectral fitting improves significantly with $\chi^2$/dof = 215/229 for the rising phase spectrum and 233/229 for the decay phase spectrum. Our best-fit model is {\tt pcfabs $\times$ tbabs $\times$ edge(1) $\times$ edge(2)$\times$[diskbb + nthcomp + reflionx + bremss + gauss]}. Fitted rising and decay phase spectra along with model components and residuals are shown in the top and middle panel of Figure \ref{spec}.

During both the rising and decay phase spectral fitting, We fix the value of {\tt tbabs} to the line-of-sight absorption column density 2.0 $\times$ 10$^{22}$ cm$^{-2}$ which is consistent with earlier works \citep{di90,pr95}. Since the normalization of disk blackbody model is unconstrained, we fixed it at 1350 assuming the disk reside at a distance of 6r$_g$ for a black hole mass of 6M$\odot$ and and the inclination angle of 70$^\circ$. During the rising and decay phase the disk temperature is found to be 0.74 $\pm$ 0.14 and 0.47 $\pm$ 0.22 keV respectively. While the Comptonizing plasma temperature predicted by {\tt nthcomp} model is similar (5.8 $\pm$ 0.1 keV), the photon powerlaw indices during rising and decay phases are 2.01 $\pm$ 0.11 and $<$1.65 respectively. Therefore, a hint of change in spectral shape is observed from the fitted parameter. This is also supported by fact that the ratio spectrum which is the ratio between the rising and decay spectral count rate as a function of photon energy, deviates significantly from 1 below 50 keV which is shown in the bottom panel of Figure \ref{spec}. While moving from the rising phase spectral fitting to the decay phase, the partial covering fraction decreases from 0.94 $\pm$ 0.02 to 0.72 $\pm$ 0.04, the normalization due the Comptonization model decreases by a factor of $\sim$3 and the normalization due the thermal Bremstrahlung model decreases from 5.85 $\pm$ 1.14 to 2.12 $\pm$ 0.99. Changes in spectral parameters with phase are highly consistent with that observed from the phase circle diagram of \citet{ko13} and the phase-dependent wind accretion geometry of the system presented in Figure 14 of \citet{ko13}.

\section{Discussion and conclusions}

In this work, using \asat{}/LAXPC observations of Cyg X-3 covering nearly one year, we determined the binary orbital period of the system to be 17253.56 $\pm$ 0.19 sec and a longer orbital period of the order of $\sim$35.8 days which may be consistent with the precessional motion of the jet. However two minima in $\chi^2$/dof plots are obtained at $\sim$35.8 days and $\sim$38.2 days (top left panel of Figure \ref{fit}) which is possibly caused by the fitting the lightcurve from different observations where one of them show very high average count rate (by a factor of $\sim$2). The fitting could be improved in future using more observations with high average count rate and the measurement degeneracy between two values of the slow periodic components could be resolved.     

It may be noted that previously using {\it EXOSAT} data, spanning over two years (1983-1985), the orbital period was measured to be 17252.52 $\pm$ 0.05 sec using cubic ephemeris fit \citep{va89}. Later, combining {\it ASCA}, {\it ROSAT}, {\it BeppoSAX} and {\it RXTE} observations, covering nearly seven years (1994-2001), \citet{si02} determined the orbital period of 17252.95 $\pm$ 0.04 sec and the rate of change in orbital period to be 7.7 $\pm$ 1.5 $\times$ 10$^{-10}$ s.s$^{-1}$. Compared to the orbital period measurement by \citep{si02}, the current binary period, measured from the present work is longer by $\sim$0.60 sec. In order to justify the increased orbital period, we calculated predicted orbital period by assuming their orbital period derivative measurement is robust. Considering its upper limit, the predicted current orbital period (2015-2016) should be 17253.43 $\pm$ 0.11 sec with the same binary orbital decay rate. This is consistent with \asat{}/LAXPC measurements of binary orbital period using data from epochs spanning over nearly one year. If this is true, then these results suggest that the binary system continues to slow-down the binary orbital motion at nearly similar rate.

During the flaring hard state, we detect the mHz quasi-periodic oscillations (QPOs) and energy-dependent time lag and rms spectra from three different observations of Cyg X-3 all of which were taken using \asat{}/LAXPC in 3.0-80.0 keV during the rising part of the orbital phase. Three QPOs during March 2016, in the frequency range 17-23 mHz are visible in 3.0-15.0 keV as well as 15.0-80.0 keV energy bands and such quasi-periodic variability of the order of $\sim$40-60 sec is also observed in the lightcurve obtained from combining three LAXPC units. Noticeably during the decay of the orbital phase, no QPOs have been observed at any frequency up to 100 mHz from three different observations. This implies that the QPO is quasi-persistent and mostly observed during the rising part of the lightcurve rather than decay part. It may be noted that the detection of such quasi-periodic features in the lightcurve is independent of the count rate statistics. Figure \ref{mark} clearly shows that no QPOs are detected during the decay phase of the lightcurve even if it has higher count rate than the rising phase where QPO is detected. Such phase-dependent trend of the occurrence of QPOs is also consistent with that observed from \xte{}/PCA (see bottom right panel of Figure \ref{pca}).

It may be noted that QPOs at $\sim$21 mHz were also observed by \citet{ko11} during the flaring hard X-ray state. Dominant Comptonization, lack of disk component in the spectra from our analysis indicate that the $\sim$20 mHz QPO detection occurs during hard state. This can be confirmed by the 15.0-50.0 keV {\it Swift}/BAT lightcurve that shows the AstroSat/LAXPC observation was taken during the flaring hard state. Therefore, 20 mHz QPO detection is consistent with earlier measurements. Not only that, in the context of phase, \citet{ko11} observed that 20 mHz QPOs occurred between phases 0.2 and 0.7. Interestingly, the QPOs discussed in \citet{va85} also appear exclusively in the phase interval 0.0-0.75 that corresponds to the rising part of the phase-folded X-ray light curve \citep{vi03}. All QPOs reported here are detected in the rising phase in the orbit-modulated lightcurve. 

\subsection{On the origin of mHz QPOs in Cyg X-3}

 The origin of mHz QPOs is not well-understood. Previously, mHz QPOs were promptly observed from two black hole X-ray binaries: GRS 1915+105 \citep{be00} and IGR J17091-3624 \citep{al11} and  once  from  H 1743-322  \citep{al12}. To explain the origin of mHz oscillations from GRS 1915+105 (also known as heartbeat oscillation), scenario like inward fluctuation propagation from a large distance in the accretion disk was invoked \citep{ne11} and to produce such oscillation, a major contribution of the disk is suggested. Between two sources, mHz QPOs are observed during the soft state in IGR J17091-3624 \citep{pa14} while 11 mHz QPO is observed during the hard/hard to hard-intermediate state transition in the low mass X-ray binary (LMXB) H 1743-322 \citep{al12}. During the soft state, disk blackbody flux dominates the spectra while during the hard state, non-thermal powerlaw like emission dominates \citep{re06}. Therefore, it is natural that the disk association with the origin of mHz QPOs would be important to consider. Unlike LMXBs, the scenario in Cyg X-3 is quite different as mHz QPOs are observed in this system which is dominated by the wind-fed accretion from the WR companion. The spectral analysis confirms the hard state nature when the QPO is detected and the disk component in not very strong. 
From our analysis, we observe that QPOs disappear during the falling part of the orbital phase and reappear during the rising orbital phase. Not only that, Table \ref{obs} shows that mHz QPOs are detected seven times only during the FHXR state. Therefore, broadband, orbital phase-resolved energy spectra during FHXR state may be useful to investigate the origin of mHz QPOs. However, we may note that apart from the FHXR state, \citet{ko11} detected mHz QPOs also during the FSXR state using the \xte{} archival data between 1997 and 2011 ($\sim$700 ksec). Hence it is possible that we detect QPOs only during FHXR state because of small sample size ($\sim$219 ksec (adding all exposures in Table \ref{obs})). Nevertheless our results show that chances of detecting mHz oscillations are maximum during the FHXR state. Therefore, the origin of QPOs may have connections with the spectral state. During the FHXR state, variable radio emission is observed \citep{ko10} from minor flaring ($\sim$ 300 mJy) to major flaring ($\sim$ 1Jy). We, therefore, cannot rule out the possible role of radio jet contributing to the origin of QPOs. In-depth analysis in this direction is beyond the scope of the present work.   

Comparing results from the energy spectral analysis in 4-60 keV during the rising and decay orbital phase with the phase circle diagram and the top panel of Figure 6 from \citet{ko13}, we find that, the the orbital phase when QPOs are detected from \asat{}/LAXPC corresponds to the binary phase in the range of 0.18-0.61 (see Table \ref{obs}) while no QPO is detected in the remaining phase intervals which mostly corresponds to the decay of binary phase. As observed from the energy spectral analysis presented here and the phase circle diagram presented in \citet{ko13}, major differences between fitted spectral parameters obtained from the rising and decay orbital phases are : (1) the covering fraction sharply decreases from 92-94\% during the rising phase to 70-73\% during the decay phase. A decrease in covering fraction $\sim$20\% implies that the source during the decay phase moves out of the region which is dominated by the wind accretion (see Figure 14 from \citet{ko13}). Therefore, the shortage of the supply of the accreting material during the decay phase of the orbital motion may cause QPO to disappear below the detection limit. Another indication comes from the increase in the normalization of the Comptonization and thermal Bremsstrahlung emission models by a factor of $\sim$3 and $\sim$2 respectively while moving from the decay to the rising phase spectra. During the rising phase, wind accretion is strongest as observed in the Figure 14 from \citet{ko13} and importantly, the optical depth is highest among all spectral states ($>$ 2.9; \citet{ko10}). Therefore, formation of a large scattering medium is inevitable. By modelling the effect of Compton scattering on timing properties in Cyg X-3, \citet{zd10} estimated the size of the scattering cloud R$_s$ as
\begin{equation}
R_s \simeq 1.5 \times 10^{10} cm / \tau_s \times ({\it f_c}/1 Hz)
\end{equation}
where $\tau_s$ is the optical depth of the scattering medium and {\it f$_c$} is the cut-off frequency. Considering the strongest QPO detection using LAXPC at $\sim$20 mHz (assuming no power in the PDS above this frequency) and the typical optical depth of $\sim$7 \citep{zd10} during the FHXR state, we estimated the size of the scattering cloud to be $\sim$1.6 $\times$ 10$^{11}$ cm which is smaller/comparable to the size of the binary separation of 3.0 $\times$ 10$^{11}$ cm \citep{sz08} in Cyg X-3. Therefore, the size of the scattering medium is appropriate enough to sustain the 20 mHz oscillations. Multiple QPOs detected during this phase may originate from the accretion of the oscillating clumpy stellar wind from the companion \citep{sz08} and the strength of QPOs may be boosted by the presence of the in-phase oscillation of the electron density of a high optical depth corona \citep{le98} formed by the accreted wind. This is also supported by the fact that at higher energy the thermal Comptonization component dominates the spectrum as observed from the top and middle panel in Figure \ref{spec} while the fractional rms of $\sim$21 mHz QPO increases with the photon energy as shown in the bottom panel of Figure \ref{lag}.       

One key result obtained from LAXPC is the time-lag at different energy bands. This is possible owing to the high efficiency of LAXPC at higher energy \citep{ya16a}. Figure \ref{lag} shows that relatively softer photons lag relatively harder photons with the time-scale up to $\sim$5 sec. Our spectral analysis shows that energy spectra during QPO detections have optically thick plasma from where Compton scatterings are taking places. Therefore, it is possible that high energy photons undergo Compton down-scattering to produce low-energy photons. Due to the optically-thick medium, soft photons may undergo larger down-scatterings within a scattering medium of the size of 10-12 sec. Therefore, with such time-scale and spectral nature, it is possible that softer photons will be delayed compared to harder photons of the order of 5 sec. Investigating further into the origin of soft-lag is presently out of scope of the present work.     

\section{Acknowledgement}  

We thank the referee for his/her constructive suggestions which improve the quality of the manuscript. We acknowledge the strong support from Indian Space Research Organization (ISRO) in various aspect of instrument building, testing, software development and mission operation during payload verification and guaranteed time observational phase. We acknowledge the support of TIFR central workshop during the fabrication and the testing of the payload.

\label{lastpage}

\end{document}